# Mobile Commerce and Applications: An Exploratory Study and Review

Khawar Hameed, Kamran Ahsan, and Weijun Yang

**Abstract**—Mobile commerce is enabling the development of additional revenue streams for organizations through the delivery of chargeable mobile services. According to the European Information Technology Observatory, the total amount of revenue generated by mobile commerce was reported to be less than £9 million in the United Kingdom in 2001. By 2005 this had, at least, doubled and more recent industry forecasts project significant global growth in this area. Mobile commerce creates a range of business opportunities and new revenue streams for businesses across industry sectors via the deployment of innovative services, applications and associated information content. This paper presents a review of mobile commerce business models and their importance for the creation of mobile commerce solutions.

**Index Terms** - Mobile Commerce, Mobile Applications, Mobile Environment, Mobility, Mobile Services.

———————————— ◆ ————————————

## 1 INTRODUCTION

THERE are many factors which drive the development of mobile commerce. For example, technology innovations – such as faster data transmission technologies, more capable mobile devices equipped with improved computing capacity, enhanced data storage, and better user-interfaces [1]. In addition, factors like the increasing penetration and diffusion of mobile phones into society and their associated affordances, and the integration of world economies have also increased the focus, need and dependency on mobility [1].

Lidderdale argues that, "for most organisations, information technology is an essential part of corporate life and is usually critical to taking or maintaining a leading place in the market" and that "the difficulty is that too often there is too large a gap between the information technology and the business' strategies" [2]. In order to address this difficulty, one approach is for organisations to establish enterprise architecture-based perspectives to fill the gap between their information technology and business strategies. For example, Intel established the Intel@ Distributed Enterprise Architecture eXtended (Intel@ DEAX) in order to help its organisations link business process models and wireless network infrastructure [3]. However, in order to understand the broader phenomenon and universe of discourse created through the coupling of mobile technologies and commerce, it is necessary to explore business models for mobile commerce first as a pre-cursor to determining how commercial entities should adapt and operate to maximize revenue streams, or evaluate and recreate their services differently to leverage the full competitive advantages of mobile commerce.

The emergence of electronic commerce has caused a revolution in the commercial environment through fundamentally altering the end-to-end process of undertaking commercial transactions through electronic means - for example, sending product orders and invoices though a network [1]. However, with the advent of wireless networks, the rapid proliferation of mobile devices in recent years, and the demand for associated value-added services, the area of mobile commerce has also emerged. This emergence is causing another revolution in the commercial environment. Mobile commerce makes business mobility a reality [2]. For example, enabling a stock investor to access the latest stock market information and undertake associated transactions using mobile devices connected to wireless network at any time and from anyplace.

With the rapid growth of mobile commerce, more and more organizations are rapidly transforming their capability to enable the delivery of mobile commerce. However, this transformation is not without difficulty, specifically in that "the structures, processes and systems that organizations have today are inflexible: they are incapable of rapid change. More computer hardware, or software, or packages, or staff, or outsourcing are not the solution." [1] Given this difficulty, one approach for organizations today is to establish an enterprise architecture to govern and accelerate the transitions to mobile commerce.

## 2 BUSINESS MODEL OF MOBILE COMMERCE

Coursaris and Hassanein from McMaster University in Canada provide an understanding of a business process model of mobile commerce. This can be seen as a detailed conceptualization of an enterprise's business activities at an m-commerce business level and also serve as a basis for the implementation of business processes in a wireless environment [4].

————————————

- *K. Hameed is with the Faculty of Computing, Engineering and Technology, Staffordshire University, Stafford, ST18 0AD, UK.*

- *K. Ahsan is with the Faculty of Computing, Engineering and Technology, Staffordshire University, Stafford, ST18 0AD, UK.*

- *Weijun Tang is with the Faculty of Computing, Engineering and Technology, Staffordshire University, Stafford, ST18 0AD, UK.*

To illustrate their point, they give a model of mobile commerce that reflects on the business process of mobile commerce and mobile consumer's possible activities. An associated business model is shown in figure 1.

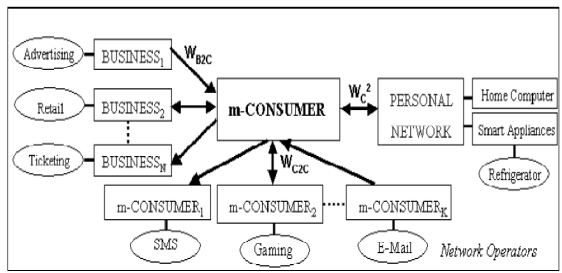

Fig. 1. Business Model for Mobile Commerce - Coursaris & Hassanein

Their three entities are Mobile Businesses, Mobile Consumers and Personal Networks.
1. **Mobile Businesses** essentially refer to service providers that a consumer may need or want wirelessly interact with for business-related purposes – e.g. for some particular service delivery [4].
2. **Mobile Consumers** refer to individuals that may need or want to wirelessly interact with service providers to procure some specific service for personal purposes [4].
3. **Personal Networks** refer to the communications infrastructure owned and accessed by con-sumer and which form part of the end-to-end transaction. This infrastructure may be specific to a particular technology or environmental context [4].

In addition, figure 1 shows that the different entity types form four kinds of business perspectives of mobile commerce.

## 2.1 Wireless Business-to-Consumer Model
The Wireless Business-to-Consumer ($W_{B2C}$) model adopts the perspective of activities, as opposed to transactions, to describe the interactions and relationships between organisations and mobile consumers [5]. These activities are focused around service delivery to mobile consumers through wireless networks, specifically including mobile advertising, mobile shopping, mobile ticketing, mobile stock trading, and mobile banking.

The WB2C model has advantages when compared with traditional face-to-face transaction models and leverages the specific benefits afforded by mobility: (1) Sales and purchases can be conducted at anytime and anyplace. (2) Personalized services are provided to consumers in a convenient and rapidly deployed manner. (3) Reduced delivery costs result in competitive commodity prices and improved efficiency of transactions.

## 2.2 Wireless Business-to-Business Model
The Wireless Business-to-Business ($W_{B2B}$) model describes commercial transactions between organisations - such as between a manufacturer and a wholesaler, or between a wholesaler and a retailer [6]. Its major characteristic is that the supply chain of constituent organisations required for the transactions, and the associated transaction process itself is predominately underpinned by wireless networks and communications. This aims to harmonise the collaborative basis of the transaction communication (again leveraging the benefits afforded by mobility) with a view to optimising the spectrum of costs associated with service delivery.

## 2.3 Wireless Consumer-to-Consumer Model
The Wireless Consumer-to-Consumer ($W_{C2C}$) model views activities occurring between consumers through some third party [6]. These activities are fairly common and typical such as E-mail, SMS, gaming, web access, and location-based activities.

## 2.4 Wireless Consumer-to-Self Model
Mobile devices are increasingly becoming part of end-to-end and sensor-based systems. For example, smart phones and personal networks can be used to communicate with or control other devices such as vehicles (in telematics scenarios), smart refrigerators, and domestic media recorders. This form of usage is encapsulated by Coursaris and Hassanein in a model called Consumer-to-Self. The Wireless Consumer-to-Self ($W_C^2$) model thus provides a perspective of the interaction of related activities that occur amongst mobile consumers themselves in context of personalised or context-based scenarios [4].

Camponovo & Pigneur from the University of Lausanne propose that a business model of mobile commerce includes a description of various business actors and their roles, a description of the potential benefits for those various business actors and, crucially, a description of the sources of revenues [7]. Figure 2 illustrates such a model of mobile commerce including the associated stakeholders.

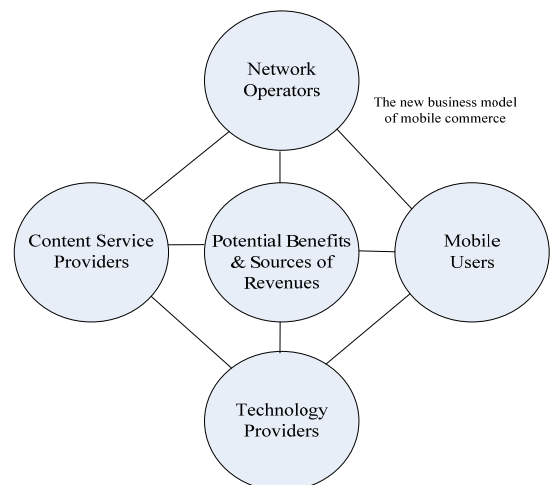

Fig. 2. A current business model of mobile commerce - Campanovo & Pigneur [7].

As shown in figure 2, Network Operators, Content Service Providers, Mobile Users, Technology Providers and potential benefits & sources of revenues comprise the essential structure and basis of interaction for the model.

## 3 CHARACTERISTICS OF VARIOUS MOBILE COMMERCE APPLICATIONS

Coursaris and Hassanein researched four different areas of business applications in mobile commerce. These four application areas included Communication, Information, Entertainment, and Commerce. According to their research, they found that mobile commerce applications had unique concerns due to the application-specific need of targeting a spectrum of different user needs. The researchers tabulated the characteristics of the mobile commerce applications (Table 1).

Table 1 Characterisation of Application (Coursaris & Hassanein)

| Business Application | Needs[1] 1 2 3 4 | Interaction Mode | User market in millions, 2005[9] | Value | Concerns | Technology Requirements[8] |
|---|---|---|---|---|---|---|
| **Communication** | | | | | | |
| - Voice | ✓ ✓ ✓ ✓ | $W_{B2C}$ $W_{C2C}$ | 1268 | Highest | Cost, Privacy | 1G / 2.5G, Voice module |
| - SMS | ✓ ✓ ✓ ✓ | $W_{B2C}$ $W_{C2C}$ | 1268 | Highest | Cost | 2G / 2.5G, WAP 2.0 |
| - e-Mail | ✓ ✓ ✓ ✓ | $W_{B2C}$ $W_{C2C}$ | 200[A] (by 2004) | Highest | Cost | 2G / 2.5G, WAP 2.0 |
| - Data Transfer | ✓ ✓ ✓ ✓ | $W_{B2C}$ $W_{C2C}$ $W_C^2$ | 2.8 (residential) 9.5 (Total) | Highest | Cost | 2.5G / 3G |
| **Information** | | | | | | |
| - Web browsing | ✓ ✓ ✓ | $W_{B2C}$ | 614 | Highest | Cost, Usability | 2G / 2.5G /3G, WAP 2.0 |
| - Traffic/Weather | ✓ | $W_{B2C}$ | N/A[6] | Highest | Privacy, Usability | 2G / 2.5G, LBS[7] |
| **Entertainment** | | | 775 (Total) | | | |
| - Gaming | ✓ ✓ | $W_{B2C}$ $W_{C2C}$ | 200[B] | Highest | Cost, Usability | 2G / 2.5G /3G, WAP 2.0 |
| - News/Sports | ✓ | $W_{B2C}$ | N/A | High | Cost, Usability, Privacy | 2G / 2.5G |
| - Downloading Music/Video/Img. | ✓ | $W_{B2C}$ | N/A | Medium | Download times, Cost | 2.5G / 3G, WAP 2.0 |
| - Horoscope/ Lottery | ✓ ✓ | $W_{B2C}$ | N/A | Low | Cost, Privacy | 2G |
| **Commerce** | | | | | | |
| - Ticketing (e.g., Event, Cinema) | ✓ | $W_{B2C}$ | N/A | Highest | Cost, Usability, Security, Privacy | 2G / 2.5G |
| - Pre-Payment | ✓ | $W_{B2C}$ | 18.3 (by 2003) | Highest | Security | 2G / 2.5G, Real-time Billing |
| - Banking | ✓ | $W_{B2C}$ | 798 | High | Security, Privacy | 2G / 2.5G |
| - Advertising | ✓ | $W_{B2C}$ | $16-23 billion | Medium | Privacy (Spam) | 2.5G/3G, LBS, WAP 2.0 |
| - Retailing | ✓ | $W_{B2C}$ | 469 | Medium | Security, Privacy, Usability | 2.5G / 3G, LBS, WAP 2.0 |

As shown in Table 1, the applications have been examined from five dimensions - Interaction Mode, User Market, Value, Concerns, and Technology Requirements.

## 4 TECHNOLOGY INFRASTRUCTURE FOR MOBILE COMMERCE

Siau *et al.* from the University of Nebraska-Lincoln in the USA found that the function of a mobile commerce infrastructure level is to enable integration and connection to the organisation's businesses and network operators [8]. They introduce two core components at m-commerce infrastructure level – these are mobile communication technology, and information exchange technology.

### 4.1 Mobile Communication Technology

According to their research, Siau *et al.*, point out that mobile communication technology is designed to transport data and information in coded digital form between various computers that support storage, retrieval, updates and processing for mobile end-users [8]. They introduce the three major mobile communication network infrastructures: GSM infrastructure, GPRS infrastructure and UMTS infrastructure. Each of these network infrastructures has impact upon the mobile end—user capability in terms of application capability, and associated information access (for example, the association of higher-bandwidth mobile communications with rich media information access and transactions.) All infrastructures can offer support for mobile commerce in varying degrees – including mobile advertising, coupon or ticketing services, and paid-for entertainment services.

### 4.2 Information Exchange Technology

Siau *et al.* state that fundamental information exchange technologies mainly include Extensible Markup Language (XML), Wireless Markup Language (WML) and Short Message Service (SMS). WML is considered as a derivation from XML having been developed especially for WAP [8]. SMS, albeit being different in nature from XML and WML still features and reference to it is no surprise given its penetration rate and capability as a core and effective vehicle to support messages as part of mobile commerce transactions.

### 4.3 Wireless Application Protocol

Constantiou & Polyzos from the Athens University of Economics and Business state that WAP is an open, global specification that empowers mobile users with wireless devices to access and interact with information services [9]. They also point out that WAP can be adopted in most mobile devices and different mobile communication systems. In addition, they refer to the WTLS specification found in WAP which implements options for authentication and encryption optimized for use in the mobile environment [9]. This potentially provides a basis for developing secure end-to-end mobile commerce transactions.

### 4.4 Wireless Middleware

Lam and Yazdani from the University of Cape Town point out the proliferation of wireless networks as being a key driver for the development of wireless middleware as a specialized subset of middleware [10]. They classify wireless middleware into 'traditional middleware' and 'more recent middleware.' Traditional middleware includes messaging middleware and distributed transaction-processing monitors. More recent middleware includes component-based middleware, and component technology XML middleware. Moreover, Lam and Yazdani found that wireless middleware is presented with many challenges due to the potentially complex nature and characteristics of wireless environments [10].

## 5 RECENT BUSINESS MODELS FOR MOBILE COMMERCE

More recent business models of mobile commerce describe the operations and processes of how organisation achieve benefit from providing mobile services and transfer of benefits across stakeholders with focus on the value chain [11]. One model of mobile commerce comprises three components including: participants in the supply chain, types of service (such as news, information, loca-tion-based services, online shopping, entertainment), and the sources of profits (such as communication fees, commission, transaction fees, subscription fees). Figure 3 describes a more recent business model of mobile commerce.

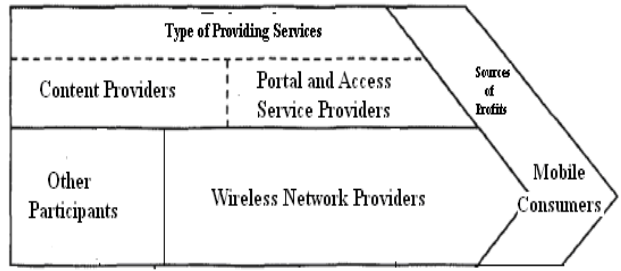

Fig. 3. A more recent business model of mobile commerce - Xiaobo & Qi [12].

This business model (Fig. 3) includes four types of participants that service the Mobile Consumers through the provision of mobile services. These services provide a revenue stream and a source of profits. The integrated stakeholder/participant approach enables all service delivery parties to exploit shared revenue opportunities from mobile consumers. The network operators, however, hold a dominant or at least a controlling position in this monetary flow [13]. Given this wider spectrum of stakeholders and service-delivery potential, the development of business models for mobile commerce is potentially a more complex issue than that for traditional business.

The integrative aspects of such an approach to business models for mobile commerce appear to be an approach that is not only appropriate, but also necessary. For example, the delivery of value-added mobile services is normally dependent on a combination of services where the final delivered service is a transparent aggregation of required components. A typical example of this is a location-based service where significant aggregation of component services and content occurs (such as aggregation of basemap content, overlays, location-determining technology, and personalised services based on user profiles, and an associated cost model for context-based information). Another example is that of the typical model of mobile advertising services which includes personalization of content based on consumer profiles, advertisers, content providers, and wireless network providers – again demonstrating the diversity of components and stakeholders, and the need to encapsulate all these.

The following sections present and review models from Yan and Yufei [11]

### 5.1 A Business Model of Mobile Information Service

In a business model of mobile information services, the key participants are Content Providers, Wireless Network Operators and Mobile users. The information services include simple services such as news, weather, stock information, as well as more complex location-based information services such as hotel location information and availability. The source of revenue is based on subscription fees paid by mobile consumers. Subscription versus transaction fees is preferable due to ease of collection and the associated predictability of these as a revenue stream [14]. Figure 4 shows a simple business model of mobile information service.

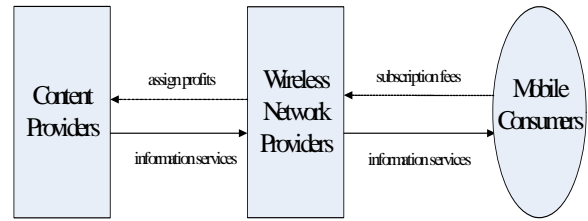

Fig. 4. The business model of mobile information service - Yan & Yufei [11]

Content providers offer a wide spectrum of information services to mobile consumers through wireless network providers. This normally requires network subscription fees in addition to fees for information services, delivered through a billing mechanism that accommodates revenue distribution and sharing.

### 5.2 A Business Model of Mobile Advertising Services

In a business model of mobile advertising, the key participants are advertisers, content providers, wireless network providers and mobile consumers, with major services focusing on advertising. The source of revenue still includes subscription fees paid by mobile consumers. Figure 5 shows a business model of mobile advertising.

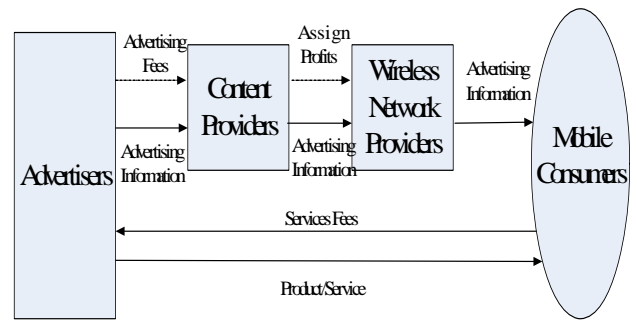

Fig. 5. The business model of mobile advertising - Yan & Yufei [11]

In this model of mobile advertising, advertisers benefit from mobile consumers who purchase commodities. Advertisers may also distribute profits to content providers and wireless network providers as part of a profit sharing agreement. In addition, wireless network providers also benefit from network subscription fees paid by mobile consumers

### 5.3 A Business Model of Mobile Office Services

A business model that governs mobile office services can help increase efficiency for mobile field workers and support anytime, anyplace access to organisations' resources. According to IDC's definition, mobile workers are not in the office for 20% or more of their time, but do need access to resources beyond the physical boundary of the organisation [11]. Figure 6 shows business model of sup-porting mobile workers.

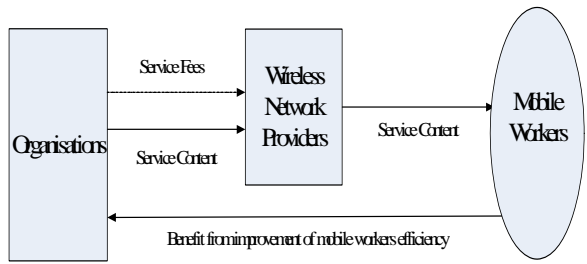

Fig. 6. The business model of supporting mobile workers - Yan & Yufei [11]

This model of mobile office services includes the organisation, wireless network providers and mobile workers. Wireless network providers offer supporting services and content to mobile workers and secure service fees from organizations. Organizations adopt a similar role to that of content service providers, offering applications and service content to mobile workers through the wireless network providers. Organizations benefit from enhanced service delivery, optimisation of services and costs, and enhanced competitive capability.

## 6 CONCLUSION

Mobile computing has permeated almost all aspects of life - across personal, social and economic systems. The maturity of the technology and its perceived and actual affordances has allowed its application to commerce – yielding the area of mobile commerce (m-commerce). The constantly evolving mobile technology landscape comprising devices, software, applications, and networks present challenges for developing mobile commerce applications. The strategic-level focus and understanding of business models for mobile commerce enables adopters to focus on developing innovative value-added solutions that exploit the commercial benefits of mobility.

**Khawar Hameed** is a Principal Lecturer in the Faculty of Computing, Engineering & Technology at Staffordshire University. His research is in the area of mobile and flexible working, enterprise architectures to support mobile information systems, and mobile learning. He has been a key driver in the adoption of mobile computing and technology within the Faculty's portfolio and has helped drive the development of undergraduate and post-graduate degrees in this technology area. He has contributed extensively to the development and delivery of externally funded projects and academic-industrial collaborations in mobile/wireless technology that aim to develop and enhance the collective intellectual capital that supports the growth of mobile and wireless systems as a discipline both within academia and in industry.

**Kamran Ahsan** has an MSc in Mobile Computer Systems from Staffordshire University and MCS (Masters in Computer Science) from the University of Karachi. Kamran is a PhD researcher and lecturer in FCET (Faculty of Computing, Engineering and Technology) and a web researcher in the Centre for Ageing and Mental Health, Staffordshire University, UK. He has published several papers and has been involved in a number of UK funded research projects including KTP, NHS Trust, and Innovation Vouchers. He is Visiting Faculty at the University of Karachi. He is a consultant to businesses in IT applications, software development and web tools. His research interests are in mobile technology applications in healthcare including knowledge management.

**Weijun Yang** is a postgraduate candidate in Computer Science in the Faculty of Computing, Engineering & Technology at Staffordshire University. His focus of work is in the area of enterprise architectures to support mobile commerce systems. Previously he worked as a technical consultant in performance and functional testing of IT systems, successfully developing and implementing performance optimization programs. His research interests are in the development of enterprise architectures and performance metrics for evaluating the function and effectiveness of mobile commerce solutions.